\documentclass{article}
\usepackage{cite}
\usepackage{url}
\usepackage{overpic}
\usepackage{amsmath, amssymb, geometry, float}
\usepackage{graphicx}
\usepackage[inline]{enumitem}
\usepackage{authblk} 
\usepackage{xcolor}  
\usepackage{hyperref}

\newcommand{\secrefplain}[1]{\hyperref[#1]{\ref*{#1} \nameref*{#1}}}

\newcommand{\myvector}[1]{{\boldsymbol{\mathbf{ #1 }}}}

\title{Efficient Computation of the Directional Extremal Boundary\\of a Union of Equal-Radius Circles}
\author{Alexander Gribov}
\affil{Esri, 380 New York Street, Redlands, CA 92373-8100, USA\\
\href{mailto:agribov@esri.com}{agribov@esri.com}}

\date{} 

\begin{document}
	
\maketitle

\begin{abstract}
	This paper focuses on computing the directional extremal boundary of a union of equal-radius circles. We introduce an efficient algorithm that accurately determines this boundary by analyzing the intersections and dominant relationships among the circles. The algorithm has time complexity of~\(O(n \log n)\).
\end{abstract}
	
\section{Introduction}

Computing the directional extremal boundary of a union of equal-radius circles involves identifying the outermost points in a specified direction that form the outer edge of the combined circle regions. This task is essential in a wide range of applications where each circle may represent a measurement, area of influence, or coverage zone.

Without loss of generality, this paper focuses on computing the upper boundary of the union of unit circles, as this can be readily extended to circles of any equal radius by applying uniform scaling and rotation. In~this paper, we present an efficient algorithm for determining the upper boundary of overlapping unit circles. The~method begins by sorting all circle centers according to their horizontal positions. Once sorted, the circles are processed from left to right to build the boundary incrementally, accounting for overlaps and overshadowing. Specifically, it identifies the exact horizontal positions where one circle's boundary overtakes or merges into another's. By maintaining a dynamic list of circles that currently define the boundary, the algorithm eliminates any circles that no longer contribute to the highest points, thereby preserving only those that shape the overall upper boundary.

One practical application of this algorithm is the precise localization of extremal points along elongated structures such as rail tracks or pipelines. In such cases, each circle may represent a measurement along the structure, and computing the circles' directional extremal boundary enables accurate identification of the outermost positions in the specified direction.

Another application arises in robotic workspace modeling, where a robotic arm moves along rails in one direction and extends in another. Here, circles can represent obstacles and safety zones: the circle centers reflect object positions, while radii model object sizes, arm dimensions, and required safety distances. The~algorithm can then determine safe regions in which the arm may operate without collisions.

\section{Problem Statement}

\begin{figure} [htb]
	\centering
	\includegraphics[scale=1.0]{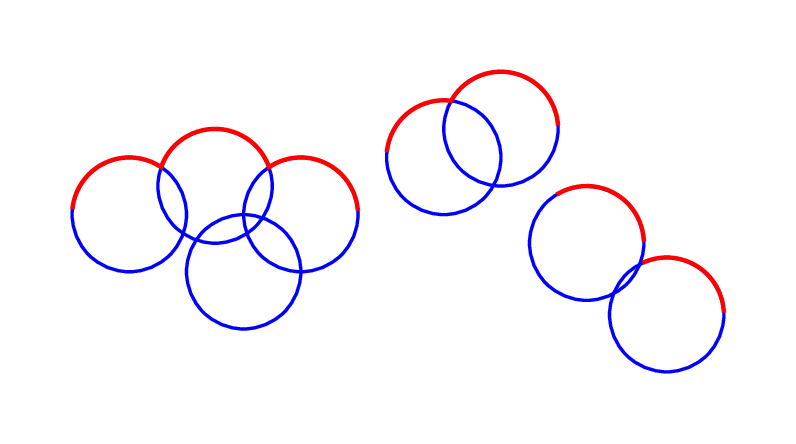}
	\caption
	{
		Visualization of the upper boundary of unit circles. The circles are shown as blue and red arcs, with the red arcs forming the upper boundary, which defines the highest points along the \(y\)-axis.
	}
	\label{fig:unit_circles}
\end{figure}

Given \( n \) unit circles \( C_i \) centered at \( \myvector{c}_i = (x_i, y_i) \) for \( i = \overline{1,n} \), our objective is to compute the maximum envelope function \( f(x) \) representing the upper boundary of their union:
\begin{equation*}
	f(x) = \max_{|x - x_i| \leq 1, \; i \in \overline{1,n}} \left( y_i + \sqrt{1 - (x - x_i)^2} \right).
\end{equation*}
This function identifies the highest point among the circles at any horizontal position \( x \), as illustrated in Figure~\ref{fig:unit_circles}.

Circles whose centers are separated by a horizontal distance of at least twice the radius (i.e., two units for unit circles) do not intersect and can be processed independently. Furthermore, if multiple circles have the same horizontal coordinate \( x_i \), only the circle with the highest vertical coordinate \( y_i \) contributes to the upper boundary.

\section{Algorithm Description}

The algorithm constructs the upper boundary by iteratively processing the circles. The resultant upper boundary is a piecewise curve composed of circular arcs. We group these arcs into \emph{segments}, where each segment is a horizontally continuous portion of the boundary. In other words, the upper boundary can be viewed as an ordered list of segments, with each segment separated horizontally.

Within each segment, the arcs are determined by the relationships between consecutive circles. To handle these relationships efficiently, the algorithm maintains a sequence of circles along with their corresponding horizontal transition positions. This sequence defines a single segment of the boundary. As the algorithm proceeds, it finalizes segments whenever it encounters a horizontal gap and then starts a new segment. To summarize, there are three cases for how the arcs form the boundary (two of which cause discontinuities—gaps and overshadowing):
\begin{itemize}
	\item When arcs are directly connected, meaning one arc ends where the next begins without any gap
	\item When one circle overshadows another within the same horizontal span, defining the upper boundary without introducing a horizontal gap, while the arcs remain in the same segment
	\item When the arcs are horizontally separated and therefore form distinct segments
\end{itemize}

The algorithm consists of the following steps:

\begin{enumerate}
	\item Sort the circles in ascending order based on their \( x_i \) coordinates.
	
	\item For each unique \( x_i \) coordinate, retain only the circle with the highest \( y_i \) coordinate and discard the others.
	
	\item Initialize an empty list of segments.
	
	\item Iterate through each circle \( C_i \) in the sorted list:
	\begin{enumerate}[label=(\alph*), ref=\theenumi(\alph*)]
		\item If there are no segments, create a new segment, add \( C_i \) to it, and proceed to the next circle.
		
		\item Determine the horizontal transition position between the last circle, \( C_j \), in the last segment and \( C_i \) using the method described in~\secrefplain{sec:horizontal_transition_position}.
		
		\item If the circles do not overlap (i.e., there is a horizontal gap), then create a new segment, add \( C_i \) to it, and proceed to the next circle.
		
		\item \label{AlgorithmRepeat} If the current segment has horizontal transition positions and the last horizontal transition position is more than or equal to the computed horizontal transition position \( x_u \):
		
		\begin{enumerate}[label=(\roman*)]
			\item Remove \( C_j \) and the last horizontal transition position from the current segment. Set \( C_j \) to be the circle currently at the end of the segment.
			
			\item Recompute the transition position between the segment's new last circle, \( C_j \), and \( C_i \). If~no intersection is found (which can happen due to floating-point roundoff errors), use the midpoint of their centers' \( x \) coordinates as the transition position.
			
			\item Repeat step~\ref{AlgorithmRepeat}.
		\end{enumerate}
		
		\item Otherwise, add the horizontal transition position \( x_u \) and \( C_i \) to the current segment.
	\end{enumerate}
\end{enumerate}

\section{Proof of Algorithm Correctness}

\subsection*{Uniqueness of Horizontal Transition}

\textbf{Lemma 1:} \emph{For any two unit circles \( C_j \) and \( C_i \) with centers \( \myvector{c}_j = (x_j, y_j) \) and \( \myvector{c}_i = (x_i, y_i) \) such that \( x_j < x_i \) and \( x_i - x_j < 2 \), there exists only one horizontal position \( x_u \) where the upper extremal boundaries of the circles transition from one circle to the other.}

\textbf{Proof of Lemma 1:}

For each unit circle, we only need to consider the upper half of the circle as it shadows the lower half. There are two cases to consider: when the upper halves of the unit circles intersect and when they do not intersect.

First, consider the case where the upper halves intersect. They can have at most one intersection point~\( x_u \). Suppose, for the sake of contradiction, that there are multiple intersection points. Due to symmetry about the midpoint between the centers of the circles, there would also be multiple intersections between the lower halves of the circles. This would result in more than two intersection points between the circles overall, which is impossible for circles. Therefore, only one intersection between the upper halves is possible, leading to \( C_j \)'s influence on the upper extremal boundary transitioning to \( C_i \) at the horizontal position \( x_u \).

Next, consider the case where the upper halves do not intersect. In this scenario, the domain overlap is only \( [x_i - 1, x_j + 1] \). Therefore, there can be only one horizontal transition. If \( y_j < y_i \), the transition occurs at \( x_i - 1 \); otherwise, it occurs at \( x_j + 1 \).

\subsection*{Correctness of the Algorithm}

Proving correctness for each segment (a horizontally continuous portion of the boundary) suffices to establish the correctness of the overall algorithm. Thus, in the proof we assume that every added circle has a transition point with its predecessor.

\textbf{Inductive Proof:} We prove the correctness by induction on the number of circles processed.

\textbf{Base Case:} With only one circle \( C_1 \), there is no ambiguity about the upper boundary.

\textbf{Inductive Step:} Assume that after processing \( k \) circles, the algorithm correctly represents the upper boundary. Now, consider adding the next circle, \( C_{k+1} \), increasing the total number of processed circles to~\( k+1 \).

The algorithm identifies the horizontal transition point \( x_u \) between \( C_{k+1} \) and the last circle, \( C_j \), currently in the sequence. By Lemma 1, only one such transition point exists. Two cases arise:

\begin{enumerate}
	\item \textbf{No Removals:} If the computed \( x_u \) exceeds the last recorded transition position in the sequence, then \( C_{k+1} \) contributes a new, higher portion of the boundary beyond \( x_u \). In other words, \( C_{k+1} \) is effectively extending the upper boundary further to the right. Since it is placed after all previously processed circles without displacing any of them, it must be higher than the current boundary beyond \( x_u \). Hence, adding \( C_{k+1} \) along with \( x_u \) maintains a correct and updated representation of the upper boundary.

	\item \textbf{Removal of Overshadowed Circles:} If \( x_u \) is less than or equal to the last recorded transition position, it indicates that \( C_{k+1} \) overshadows the previously last circle \( C_j \) in the current upper boundary. To maintain correctness, the algorithm removes \( C_j \) and the last transition position from the sequence and checks again with the preceding circle. By iteratively doing so, it eliminates any circle whose contribution no longer defines the highest reachable boundary segment once \( C_{k+1} \) is considered. Ultimately, when the removal process ceases, \( C_{k+1} \) is added along with a valid transition position that correctly reflects the new contributor to the upper boundary beyond that point.
\end{enumerate}

Through this inductive argument, we conclude that each step preserves correctness. Therefore, by induction, the entire algorithm correctly computes the upper boundary of the union of circles.
	
\section{Complexity Analysis}

The algorithm operates in two primary phases: sorting and processing.

\begin{itemize}
	\item \textbf{Sorting}: Sorting the \(n\) circles has \(O(n \log n)\) complexity.\cite{Knuth}
	
	\item \textbf{Processing}: Iterating through the sorted list and maintaining the current sequence of contributing circles has linear complexity, \(O(n)\),  because each circle is added to and potentially removed from the current sequence only once.
\end{itemize}

Consequently, the overall time complexity of the algorithm is \(O(n \log n)\).

\section{Binary Search for Finding the Upper Value}

In the final boundary representation, each segment corresponds to a sequence of circles. For a given segment, let \(x_{\mathrm{left}}\) and \(x_{\mathrm{right}}\) be, respectively, the smallest and largest \(x\) coordinates of the circle centers. The segment's horizontal domain is then \(\left( x_{\mathrm{left}}-1,\; x_{\mathrm{right}}+1 \right)\).\newline

To evaluate the upper boundary at specific horizontal position \(x\), proceed as follows:
\begin{enumerate}
	\item Perform a binary search across all segments to find one whose domain contains \(x\).
	\item Determine whether there is a segment containing \(x\). If no segment's domain contains \(x\), the upper boundary does not exist at that position; otherwise, proceed to the next step.
	\item Within the segment, perform a binary search on the segment's transition positions to identify the specific circle \(C_i\) that defines the boundary at \(x\).
	\item Compute the \(y\) coordinate using the circle's equation:
	\[
	y =
	\begin{cases}
		y_i + \sqrt{1 - (x - x_i)^2}, & \text{if } \lvert x - x_i \rvert < 1;\\
		y_i, & \text{otherwise}.
	\end{cases}
	\]
	Note that the case when $ 1 < \lvert x - x_i \rvert $ can potentially occur due to roundoff error.
\end{enumerate}
	
\section{Conclusion}

In this paper, we introduced an efficient algorithm for computing the directional extremal boundary of a union of equal-radius circles.\footnote{While the algorithm is presented for equal-radius circles, its core ideas can be adapted to handle ellipses that are of equal size and aligned with the specified direction.} The proposed method runs in \(O(n \log n)\) time complexity by sorting the circle centers and incrementally constructing the upper boundary. Although the sorting step can be parallelized in a straightforward manner, the described boundary-processing step is currently sequential. Nonetheless, a parallel implementation remains feasible by recursively merging partially constructed solutions derived from individual circles. In this merging phase, the transition positions from both solutions are processed in a manner analogous to the sequential insertion of a new circle, ensuring that the merged solution is maintained correctly while leveraging the divide-and-conquer paradigm~\cite{Knuth}. It is important to note that the parallel implementation does not result in a segmentation of the final solution.

\section*{Acknowledgments}

The author would like to thank Lois Stuart for copy editing this paper; Manoj Lnu, principal software engineer on the ArcGIS $3$D Analyst team at Esri; and Sergey Tolstov, senior principal software development engineer on the geometry team at Esri, for providing valuable suggestions to help improve this paper as well as identifying potential areas of application.

ChatGPT was used to assist with proofreading and figure generation, supporting text refinement and enhancing the visual presentation.\cite{openaichatgpt}

\newcounter{CurrentSectionValue}
\setcounter{CurrentSectionValue}{\value{section}}
\setcounter{section}{0}

\iftrue
\renewcommand{\thesection}{Appendix:}
\else
\renewcommand{\thesection}{Appendix \Roman{section}:}
\fi

\section{Horizontal Transition Position of Two Unit Circles}
\label{sec:horizontal_transition_position}

\begin{figure}[htb]
	\centering
	\begin{minipage}{0.45\textwidth}
		\begin{overpic}[scale=0.7]{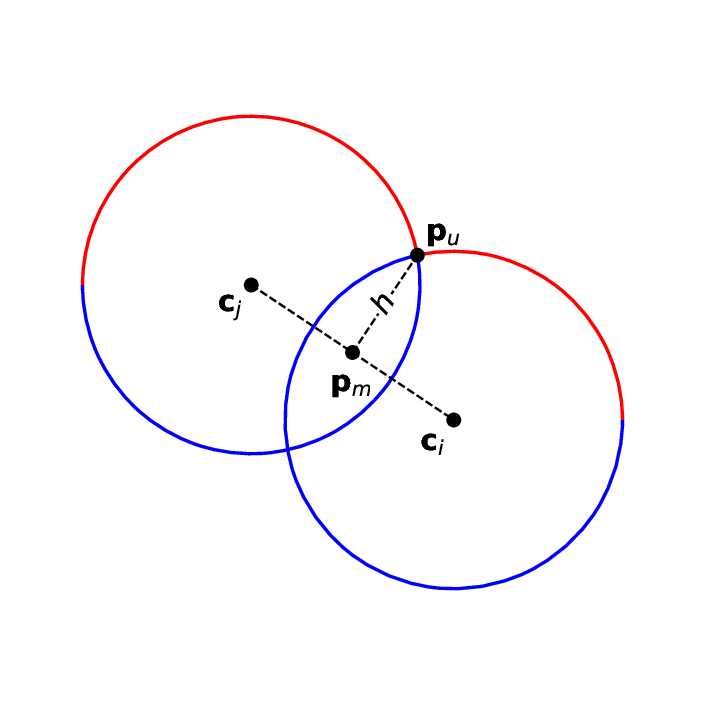}
			\put(10,10){\textbf{\ref*{item:IntersectionOfTwoCirclesA}}}
		\end{overpic}
	\end{minipage}
	\hspace{0.05\textwidth}
	\begin{minipage}{0.45\textwidth}
		\begin{overpic}[scale=0.7]{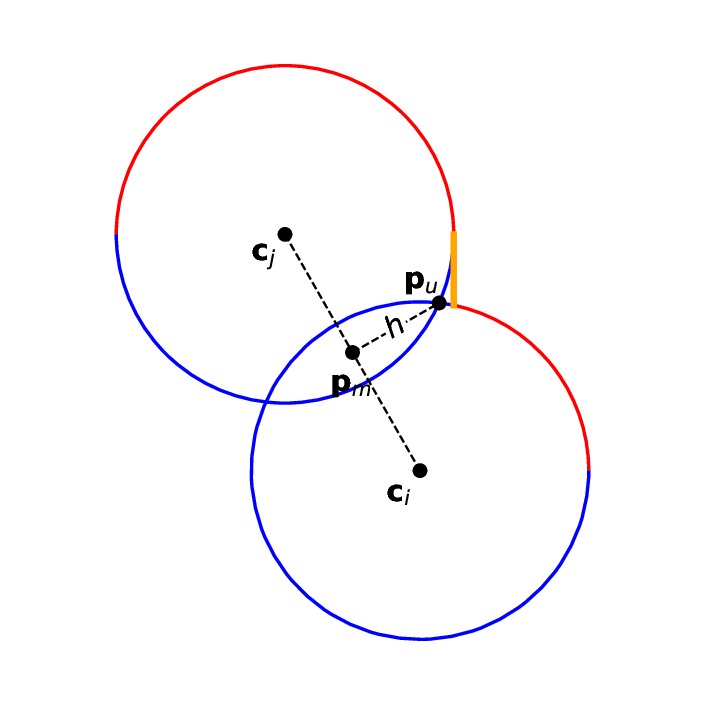}
			\put(10,10){\textbf{\ref*{item:IntersectionOfTwoCirclesB}}}
		\end{overpic}
	\end{minipage}
	\caption[]
	{
		Comparison of the upper boundaries of two unit circles.
		\begin{enumerate*}[label=(\textit{\Alph*}), ref=\textit{\Alph*}]
			\item \label{item:IntersectionOfTwoCirclesA} Intersecting circles with a continuous upper boundary.
			\item \label{item:IntersectionOfTwoCirclesB} A discontinuity (in orange) when one circle overshadows the other.
		\end{enumerate*}
	}
	\label{fig:intersection_two_circles}
\end{figure}

Consider the two unit circles \( C_j \) and \( C_i \) with centers at \( \myvector{c_j} = (x_j, y_j) \) and \( \myvector{c_i} = (x_i, y_i) \), respectively, where we assume \( x_j < x_i \), as shown in Figure~\ref{fig:intersection_two_circles}.

If \( x_j + 2 \leq x_i \), the circles do not overlap horizontally. Even the edge case \( x_j + 2 = x_i \) indicates tangential contact, treated here as nonintersecting.

Compute the distance between the centers:
\begin{equation*}
	d = \sqrt{(x_i - x_j)^2 + (y_i - y_j)^2}.
\end{equation*}

If the circles intersect and their centers are not too close (i.e., \( \epsilon < d < 2 \), where \( \epsilon \) is a small positive number accounting for floating-point precision), we calculate their upper intersection point \( (x_u, y_u) \) as follows:

\begin{itemize}
	\item Compute the midpoint between the centers:
	\begin{equation*}
		\myvector{p_m} = \frac{1}{2} \left( \myvector{c_j} + \myvector{c_i} \right).
	\end{equation*}
	\item Compute the unit vector \( \myvector{p_n} \) perpendicular to the line connecting the centers and pointing upright:
	\begin{equation*}
		\myvector{p_n} = \frac{1}{d} \left( - (y_i - y_j),\ x_i - x_j \right).
	\end{equation*}
	\item Calculate the distance \( h \) from the midpoint to the intersection points:
	\begin{equation*}
		h = \sqrt{1 - \left( \frac{d}{2} \right)^2 }.
	\end{equation*}
	\item Compute the upper intersection point:
	\begin{equation*}
		\myvector{p_u} = \myvector{p_m} + h \cdot \myvector{p_n},
	\end{equation*}
	where \( \myvector{p_u} = (x_u, y_u) \).
\end{itemize}

If \( \max(y_j, y_i) < y_u \) (see Figure~\ref{fig:intersection_two_circles}\ref*{item:IntersectionOfTwoCirclesA}), then the horizontal transition position is \( x_u \). 
Otherwise, referring to Figure~\ref{fig:intersection_two_circles}\ref*{item:IntersectionOfTwoCirclesB}, the horizontal transition position is determined by:
\begin{equation*}
	\begin{cases} 
		x_i - 1, & \text{if } y_j < y_i, \\ 
		x_j + 1, & \text{otherwise}.
	\end{cases}
\end{equation*}

\bibliographystyle{plainurl}
\bibliography{UnionCirclesBoundary}

\end{document}